\documentclass[journal,12pt,draftclsnofoot,onecolumn]{IEEEtran}

\usepackage{amsmath,amssymb,epsfig,cite,footnote,authblk}
\usepackage{color,hyperref}

\hypersetup{
    colorlinks,%
    citecolor=blue,%
    filecolor=blue,%
    linkcolor=blue,%
    urlcolor=blue
}
\definecolor{bluecolor}{rgb}{0,0.,1.}

\definecolor{redcolor}{rgb}{.7,0.,0.}

\newcommand{\es}[1]{\begin{equation}\begin{split}#1\end{split}\end{equation}}

\newcommand{\R}{\mathbb{R}}

\newcommand{\E}{\mathbb{E}}

\newcommand{\dd}{\textrm{d}}

\newcommand{\Var}{\mathrm{Var}}
\newcommand{\Cov}{\mathrm{Cov}}

\begin{document}

\title{Counting Geodesic Paths in 1D VANETs}
\author[1]{Georgie Knight}
\author[1]{Alexander P. Kartun-Giles}
\author[2]{Orestis Georgiou}
\author[1]{Carl P. Dettmann}
\affil[1]{School of Mathematics, University of Bristol, University Walk, Bristol, BS8 1TW, UK \thanks{g.knight@bristol.ac.uk}}
\affil[2]{Toshiba Telecommunications Research Laboratory, 32 Queens Square, Bristol, BS1 4ND, UK}
\maketitle

\begin{abstract}
In the IEEE 802.11p standard addressing vehicular communications, Basic Safety Messages (BSMs) can be bundled together and relayed as to increase the effective communication range of transmitting vehicles. 
This process forms a vehicular ad hoc network (VANET) for the dissemination of safety information. 
The number of ``shortest multihop paths" (or geodesics) connecting two network nodes is an important statistic which can be used to enhance throughput, validate threat events, protect against collusion attacks, infer location information, and also limit redundant broadcasts thus reducing interference. 
To this end, we analytically calculate for the first time the mean and variance of the number of geodesics in 1D VANETs.
\end{abstract}

\section{Introduction \label{sec:intro}}

Vehicular ad-hoc networks (VANETs) are formed by a collection of vehicles, wirelessly connected to each other to form a communication network.
These networks are typically one dimensional as they dynamically and rapidly self-organise on roads and highways, although communication with road side access point infrastructure is possible under different scenarios. 
Primarily, vehicle to vehicle (V2V) and vehicle to infrastructure (V2I) communications involve safety related issues, such as collision warnings aimed at preventing imminent car accidents through broadcasting and relaying messages, thereby increasing local situation awareness \cite{hafeez2013performance}. 
V2V communications can also be exploited for applications such as intelligent cruise control or platooning, traffic information and management, as well as internet access and advertising. 

The IEEE 802.11p standard defines a wireless area network (WLAN) for dedicated short range communication (DSRC) among vehicles.
The standard defines protocols for the physical and MAC layers, has a 75 MHz bandwidth allocated at 5.9 GHz, and is the prime candidate currently being deployed in order to get IEEE 802.11p equipped cars on the roads \cite{kenney2011dedicated}.
Under the standard, it is possible to bundle together information on position, speed, direction, brake information, steering wheel angle, threat-events, etc., and append them to a \textit{basic safety message} (BSM) which is then  broadcasted \cite{najafzadeh2014bsm}. 
Vehicles within range can then actuate on this information, edit it, or append to the content message, and re-broadcast, thus locally flooding the network.

\begin{figure}[t]
\begin{centering}
	\includegraphics[scale=0.6]{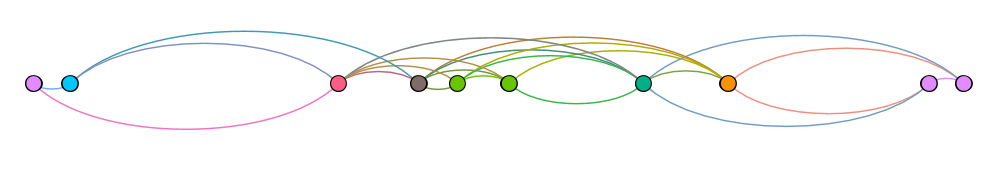}
\par\end{centering}
\caption{A one-dimensional unit disk graph ($r_0=1$) used to model a VANET. 
The geodesic length between the two extreme nodes is $k=3$.}
\label{fig:graph}
\end{figure}

Flooding algorithms are commonplace in ad hoc networks, however here the algorithm is also spatially constrained to run along a one-dimensional road network. 
Such networks are typically modelled as random geometric graphs \cite{penrose2003random} formed by a 1D Poisson point process (PPP) and a communication range $r_0$ (see Fig.\ref{fig:graph}) directly related to SNR \cite{georgiou2013connectivity} thus lending themselves to mathematical analysis and engineering.
A major challenge in VANETs is the timeliness and latency in which information must arrive to be useful to a fast approaching vehicle.
Hop-count statistics find application in a variety of other settings, e.g., in gas pipe sensor networks \cite{stoianov2007}, nanowires \cite{larsen2015}, and map navigation problems in general.
Therefore, hop-count statistics have been extensively studied in 1D  \cite{vural2007} and 2D networks \cite{mao2010,chandler1989}.
They were first studied by Chandler \cite{chandler1989}, who looked at the probability that two wireless network nodes can communicate in $k$ hops.
Such information can further assist the calculation of network centrality measures \cite{giles2015}, or achieve range-free localisation \cite{nguyen2015}.

In this paper we are concerned with the statistical properties of the \textit{shortest} multihop paths, also referred to as \textit{geodesics}, between nodes in 1D VANETs.
To this end, we calculate for the first time the first few moments of the number of geodesics $\sigma$ between nodes in a 1D VANET, as a function of the Euclidean distance $L$ between them and the vehicle density $\lambda$.
Clearly for $L\!\in\!((k\!-\!1)r_0, kr_0)$, the shortest possible path is of length $k$ hops, employing just $k-1$ relay nodes, thus defining a fundamental upper limit on the latency involved with such transmissions.
On the other hand, due to the broadcast nature of wireless transmissions, multiple  BSMs containing similar information may arrive via different $k$-hop paths almost simultaneously.
It is therefore of interest to understand the statistical properties of the number of $k$-hop paths $\sigma_{k}$, as a function of  $r_0$, $L$, and $\lambda$.
Such statistics can be used to enhance throughput \cite{rossi2015density}, validate threat events, protect against collusion attacks, infer location information, and also limit redundant broadcasts thus reducing interference.



\section{System Model}

\begin{figure}[t]
\begin{centering}
	\includegraphics[scale=0.25]{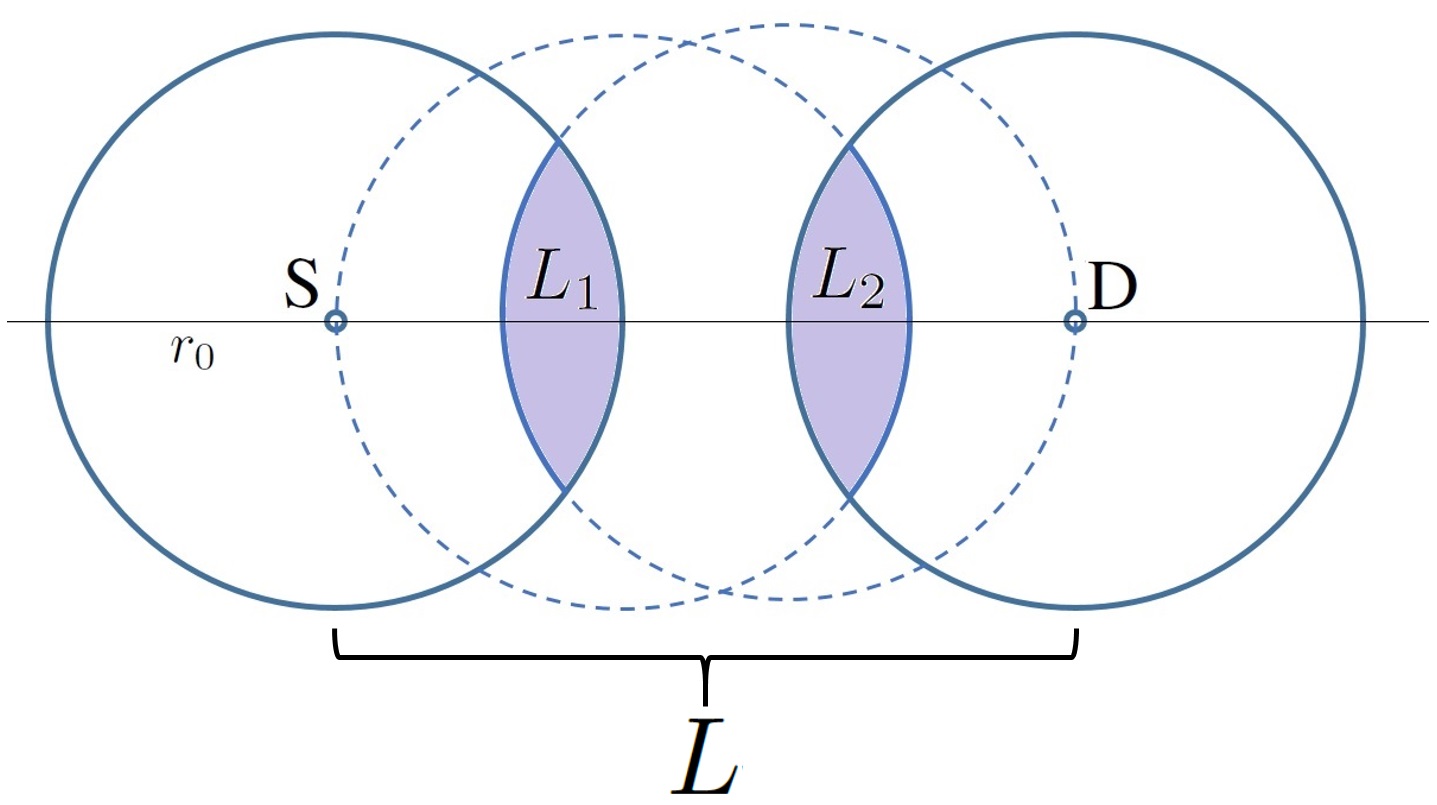}
\par\end{centering}
\caption{System Model: All 3-hop paths between source (S) and destination (D) nodes separated by a distance $L\in(2r_0, 3r_0)$ must involve at least 1 relay node located in each of the shaded  ``lenses".}
\label{fig1}
\end{figure}

\begin{figure}[t]
\begin{centering}
	\includegraphics[scale=0.3]{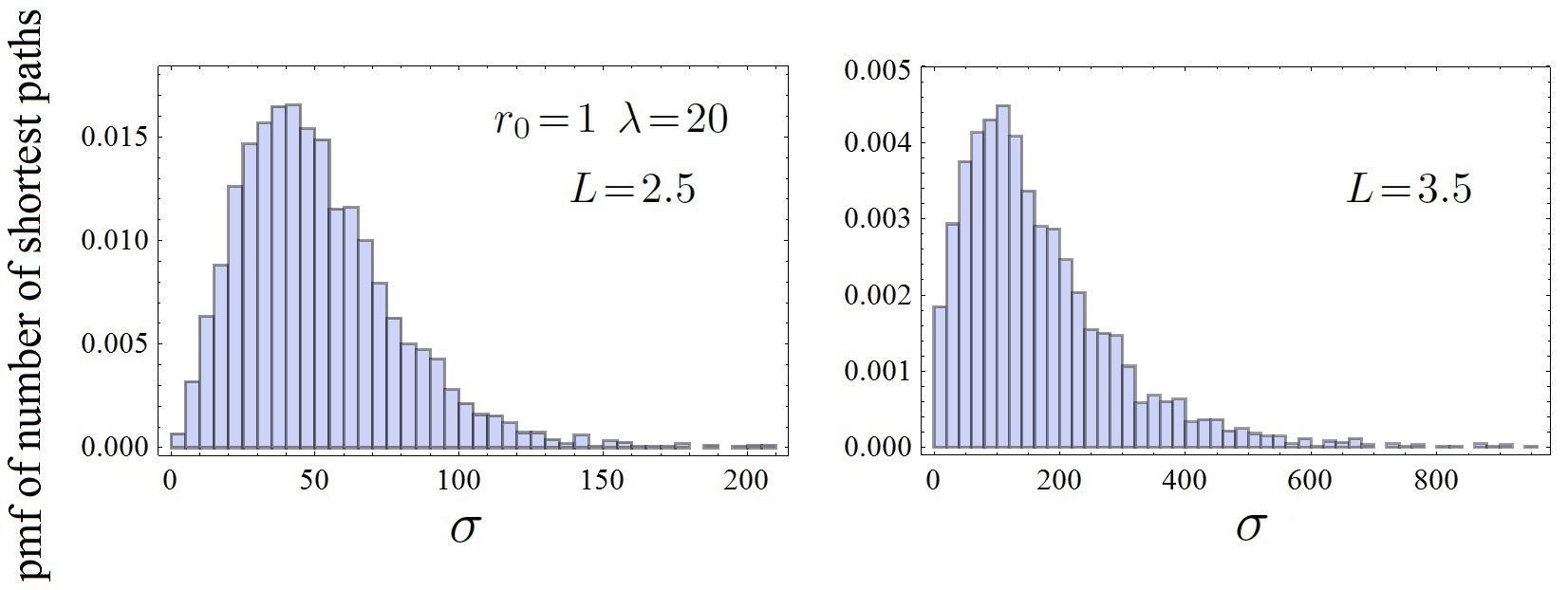}
\par\end{centering}
\caption{Probability mass function of the shortest paths $\sigma$ for $r_0\!=\!1, \lambda\!=\!20,$ and $L\!=\!2.5, 3.5,$ such that geodesics are of length $3$, and $4$ respectively.}
\label{fig2}
\end{figure}

Consider a source node $\mathcal{S}$ located at the origin, and a destination node $\mathcal{D}$ a distance $L\!>\!2r_0$ to the right of $\mathcal{S}$ along the positive real line.
Further, consider a 1D PPP of density $\lambda$ vehicles per unit length forming on the real line, with each point (node) representing a vehicle along an infinite stretch of road.
Nodes are then connected via communication links whenever their Euclidean distance is less than a predefined communication range $r_0$ (see Fig. \ref{fig1}), thus forming a 1D network.
The source and destination nodes are unable to communicate directly and must employ multihop communications in order to share information. 
Depending on the density of vehicles $\lambda$, there may be none, one, or several multihop paths connecting $\mathcal{S}$ and $\mathcal{D}$. 
The \textit{length} of these paths is the number of hops required for a message to pass between the two vehicles. 
It follows that the length of the \textit{shortest} multihop paths is $k\!=\! \lceil \frac{L}{r_0} \rceil$.
Therefore, paths of length $k$ are geodesic.
Running a breadth-first search (BFS) algorithm can find all geodesic paths in linear-time since the underlying graph is neither directed, nor weighted.
Let the set of all geodesics be described by $\Sigma(r_0,L,\lambda)$. 
Then the number of geodesic paths is
\es{
\sigma_{k} :=\text{card}\Big[\Sigma(r_0,L,\lambda)\Big]
.}
Monte Carlo simulations of the pmf of $\sigma$ are shown in Fig. \ref{fig2}.
We will first demonstrate the difficulties with obtaining the distribution of $\sigma_{k}$ for the case of $k\!=\!3$, and then calculate its first few moments for the general case of $k\!\geq\!3$.
The cases of $k\!=\!1$ and $k\!=\!2$ are trivial and therefore omitted.

\section{Distribution of Geodesic Paths}
\begin{figure}[t]
\begin{centering}
	\includegraphics[scale=0.35]{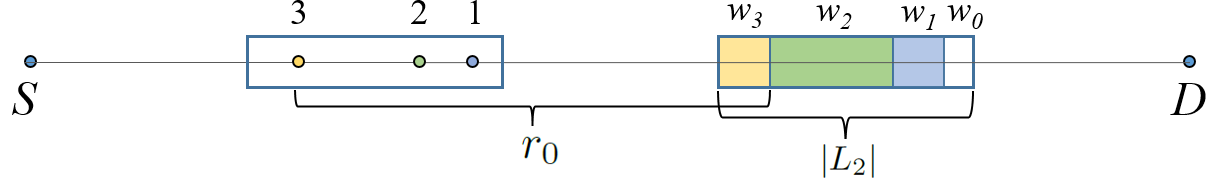}
\par\end{centering}
\caption{Schematic showing $N_1\!=\! 3$ nodes in the left lens $L_1$, and the corresponding sub-domains $w_i$ in the right lens $L_2$. Note that $w_3$ is within range from all three nodes and therefore a fourth node located in $w_3$ will connect to all in $L_1$, to form three $3$-hop paths from $\mathcal{S}$ to $\mathcal{D}$.
In contrast, $w_2$ is in range of nodes 1 and 2 (not 3), $w_1$ is only in range of node 1 (not 2 or 3), and $w_0$ is not in range from any of the nodes in $L_1$.
}\label{fig:windows}
\end{figure}

Let $L\!\in\!(2r_0,3r_0)$ and $k\!=\!3$ as in Fig. \ref{fig1} such that there are two sub-domains $L_1$ and $L_2$ within which relay nodes must be situated in order for a 3-hop path to exist.
We call these sub-domains lenses, since in two dimensions they are formed by the intersection of two equal disks.
This is because the first relay node located at a maximum distance of $r_0$ can form a 3-hop path by connecting with any node in $L_2\!=\![L\!-\!r_0,2r_0]$.
By symmetry $L_1\!=\![L\!-\!2r_0,r_0]$ such that the two lenses are of equal widths $|L_1|\!=\!|L_2|\!=\!3r_0\!-\!L$.
The number $N_1$ of relay nodes in $L_1$ is therefore a Poisson random variable with mean $\Lambda_3$, where we have defined $\Lambda_k= \lambda(k r_0 \!-\! L)$.
Moreover, for each relay node in $L_1$ there corresponds a subset of $L_2$ within which a second relay node must be located as to form a 3-hop path from $\mathcal{S}$ to $\mathcal{D}$.
Labelling the $N_1$ relays in descending distances $d_i$ from the source (i.e., $L\!-\!2r_0 \leq d_{N_1}\leq d_{N_1 \!-\!1} \leq \ldots \leq d_{1} \leq r_0$) we can identify subsets $[L-r_0,d_i+r_0]\subseteq L_2$ within which if located a second relay can successfully form a 3-hop path.
Defining the sub-domains 
$
w_i=[d_{i+1}+r_0,d_{i}+r_0]$, for $i=0,1,\ldots N_1
$
with $d_{0}=2r_0$ and $d_{N_1+1}=L-r_0$ it can be seen that a relay node in $w_i$ connects to $i$ relays in $L_1$.
We therefore arrive at a simple expression for the number of shortest 3-hop paths 
\es{
\sigma_{3}= \sum_{i=1}^{N_1}  i n_i
\label{sigma}}
where the $n_i$ is the number of relays in $w_i$ and are thus Poisson random variables with mean $\lambda w_i$ (see Fig. \ref{fig:windows}).
The widths $w_i$ are also random variables however must satisfy the constraint that $\sum_{i=0}^{N_1} w_i \!=\! 3r_0 \!-\!L$, i.e., the $n_i$ are correlated.

The pmf of $\sigma$ can be expressed as follows:
\es{
\mathbb{P} [\sigma_{3} \!=\! x ] 
\!=\! \mathbb{E}_{N_1,\mathbf{w}}\Big[ \mathbb{P} [\sigma_{3} \!=\! x  \Big|  N_1 , \mathbf{w} ] \Big]
}
where $\mathbf{w}=\{w_0,w_1,\ldots w_{N_1} \}$ and any configuration of widths $\mathbf{w}$ is equally likely.
We can attempt to obtain the pmf of $\sigma_{3}$ through the use of probability generating functions (PGFs).
Namely, we have that the PGF of the random variable $Z_i \!=\! i n_i$ is given by $G_{Z_i}(z)=\mathbb{E}[z^{i n_i}]\!=\!G_{n_i}(z^i)\!=\!e^{\lambda w_i (z^i -1)}$ since $n_i$ is Poisson distributed with mean $\lambda w_i$.
It follows that the sum of $N_1$ such random variables has a PGF given by 
\es{
G_{\sigma_{3}}(z)= \prod_{i=1}^{N_1} e^{-\lambda w_i (1-z^i)}
}
and the corresponding pmf given by 
\begin{multline}\label{general}
\mathbb{P} [\sigma_{3} \!=\! x   ] =\!  \\ \sum_{k=0}^\infty \mathbb{P}[N_i\!=\!k] \! \int_{[0,3r_0-L]^{N_1}} \!\!\!\!\!\!  \mathbf{1}(\mathbf{w})\frac{c}{k!} \frac{ \dd^k G_{\sigma_{3}}}{\dd z^k}\Big|_{z=0}  \dd w_1 \ldots \dd w_{N_1}
\end{multline}
where $\mathbb{P}[N_i=k]= \frac{\Lambda_3^k}{k!}e^{-\Lambda_3}$ and $\mathbf{1}(\mathbf{w}) $ is the indicator function equal to $1$ whenever $\|\mathbf{w}\|_1 =3 r_0 -L$ and zero otherwise such that $\int \mathbf{1}(\mathbf{w}) \dd w_1 \ldots \dd w_{N_1} = 1/c$, and $c\!>\!0$ is some normalisation constant.
Geometrically, the indicator function defines a simplex polytope with $N_1+1$ vertices at $\{\mathbf{v}_0 ,\ldots \mathbf{v}_{N_1}\}^T=(3r_0-L)\mathbf{I}_{N_1+1}$.
The integral is therefore over the surface of the $N_1$-simplex.
Recall that the $N_1$-simplex is a triangle, a tetrahedron, a 5-cell, for $N_1=2, 3,$ and $4$ respectively, and therefore is an evermore complex polytope embedded in the positive hyperoctant of $\R^{N_1+1}$ for which the integration of \eqref{general} becomes intractable.
For this reason we next restrict our study to the mean and variance of $\sigma_{3}$. 


\section{3-Hop Mean and Variance}
\label{Sec:MeanVariance}
We now describe a method which allows us to analytically derive the moments of $\sigma_3$. This involves dividing up the lenses into many small parts and making a simplifying approximation about the interactions. This allows us to treat the problem as one involving many independent random variables rather than trying to account for dependence. The final step is to take the limit of the number of divisions of the lenses to infinity, in which our approximation becomes exact.   
  
We firstly split the lenses $L_i$ into  a large number $l\!\gg\! 1$ of equally sized, disjoint domains $L_{ij}$ where $|L_{ij}|\!=\!(3r_0-L)/l$ and $ L_i\!=\!\bigcup_{j=1}^l L_{ij}$. The number of  relay nodes in each $L_{ij}$ is then a Poisson distributed random variable $Y_{ij}$ with mean $\Lambda_3/l$. For finite $l$ we make the approximation that all relay nodes in $L_{11}$ connect with all those in $L_{21}$, all those in $L_{12}$ connect with all in $L_{21}$ and $L_{22}$ etc. 
The number of shortest 3-hop paths is then given by
\es{
\sigma_{3}=\lim_{l \rightarrow \infty}\sum_{q=1}^{l} \sum_{r=1}^{q}Y_{1q}Y_{2r} 
\label{sigma2}
}
Using the independence of the $Y_{ij}$ we calculate the mean
\es{
\E[\sigma_{3}]&=\lim_{l \rightarrow \infty}\sum_{q=1}^{l} \sum_{r=1}^{q}\E[Y_{1q}]\E[Y_{2r}]\\
&= \lim_{l \rightarrow \infty}\left(\frac{\Lambda_3^2}{l^2}\right)\left(\frac{l^2+l}{2}\right)=\frac{\Lambda_3^2}{2}
\label{Expectation2}
}
To extract the variance we first define the random variable
\es{
T_q=Y_{1q}\sum_{r=1}^{q}Y_{2r} 
\label{Tq}
}
Given that the variance of a sum is equal to the sum of the variances plus the covariances we have
\es{
\Var(\sigma_{3})=\lim_{l \rightarrow \infty}\sum_{q=1}^{l}\Var(T_q) + 2\sum_{t=2}^{l}\sum_{s=1}^{t-1}\Cov(T_s, T_t)
\label{Var}
}
We first evaluate the variance of $T_q$. We use the independence of $Y_{1q}$ and $Y_{2r}$ and note that $\sum_{r=1}^{q}Y_{2,r}$ is a Poisson random variable with mean $q\lambda(3r_0 \!-\! L)/l$. 
In addition we use the mean of the square of a Poisson random variable with mean $x$ is equal to $x^2 +x$ and derive 
\es{
\Var(T_q)=\frac{q^2 \Lambda_3^3}{l^3}+q\left(\frac{\Lambda_3^3}{l^3}+\frac{\Lambda^2}{l^2}\right)
\label{VarTq}
}
Using (\ref{VarTq}) we evaluate the limit of the first sum in (\ref{Var}) as
\es{
      \lim_{l \rightarrow \infty}\sum_{q=1}^{l}\Var(T_q) =\frac{\Lambda^3}{3}+\frac{\Lambda_3^2}{2}
\label{VarTqSumLimit}
}
For the covariance terms in Eq.(\ref{Var}) we let $s<t\leq l$ and use the relation $\Cov(T_s, T_t)=\E[T_s T_t]-\E[T_s]\E[T_t]$. 
The expectation of $T_s$ is given by
\es{
      \E[T_s]=\frac{s\Lambda_3^2}{l^2}
\label{ExpTs}
}
For the expectation of the product we have via (\ref{Tq})
\es{
      \E[T_sT_t]=\E\Big[\Big(Y_{1s}\sum_{r=1}^{s}Y_{2r}\Big) Y_{1t}\Big( \sum_{r=1}^{s}Y_{2r}+\sum_{r=s+1}^{t}Y_{2r} \Big)\Big]      
\label{Eq:ExProd}
}
where by splitting the sum in $T_t$ we can factorise using the mutual independence of the terms as 
\es{
\E[T_sT_t] \!=\! \E[ Y_{1s}] \E[Y_{1t}]\Big( \E\Big[ \big( \sum_{r=1}^{s}\!Y_{2r} \!\big)^2\Big] \!+\! \E \Big[ \sum_{r=1}^{s}\!Y_{2r}\Big]\! \E\Big[\!\!\sum_{r=s+1}^{t} \!\!\! Y_{2r}\Big]  \Big)
\label{Eq:ExProd2}
}
evaluating the individual expectations and combining we have
\es{
\E[T_sT_t]=\frac{\Lambda_3^4st}{l^4}+\frac{\Lambda_3^3 s}{l^3}
\label{Eq:ExProd3}
}
Combining (\ref{ExpTs}) and (\ref{Eq:ExProd3}) we have that 
$
\Cov(T_s, T_t)=\frac{\Lambda_3^3 s}{l^3}
$.
We can now evaluate the sum of covariances in (\ref{Var})
\es{
  \sum_{s \neq t}\Cov(T_s, T_t) = \frac{2\Lambda_3^3}{l^3}\sum_{s=1}^{l-1}\sum_{t=s+1}^l \! s = \frac{2\Lambda_3^3}{l^3}\left(   \frac{l^3-l}{6} \right)
\label{Eq:VarSum}
}
Taking the limit 
$
\lim_{l \rightarrow \infty}\sum_{s \neq t}\Cov(T_s, T_t) \!=\!\frac{\Lambda_3^3}{3}
$ and combining it with (\ref{VarTqSumLimit}) we may extract the variance
\es{
\Var(\sigma_{3})=\frac{2\Lambda_3^3}{3}+\frac{\Lambda_3^2}{2}
\label{VarianceSigma}
}
Similarly we can extract higher order moments of the distribution using this technique. For example the third moment
$
\E[(\sigma_{3}-\E[\sigma_{3}])^3]\!=\! - \frac{5\Lambda_3^5}{6}-\frac{\Lambda_3^4}{5}
$, 
which can be used to analyse the skewness of the distribution.

\begin{figure}[t]
\centering
\includegraphics[scale=0.55]{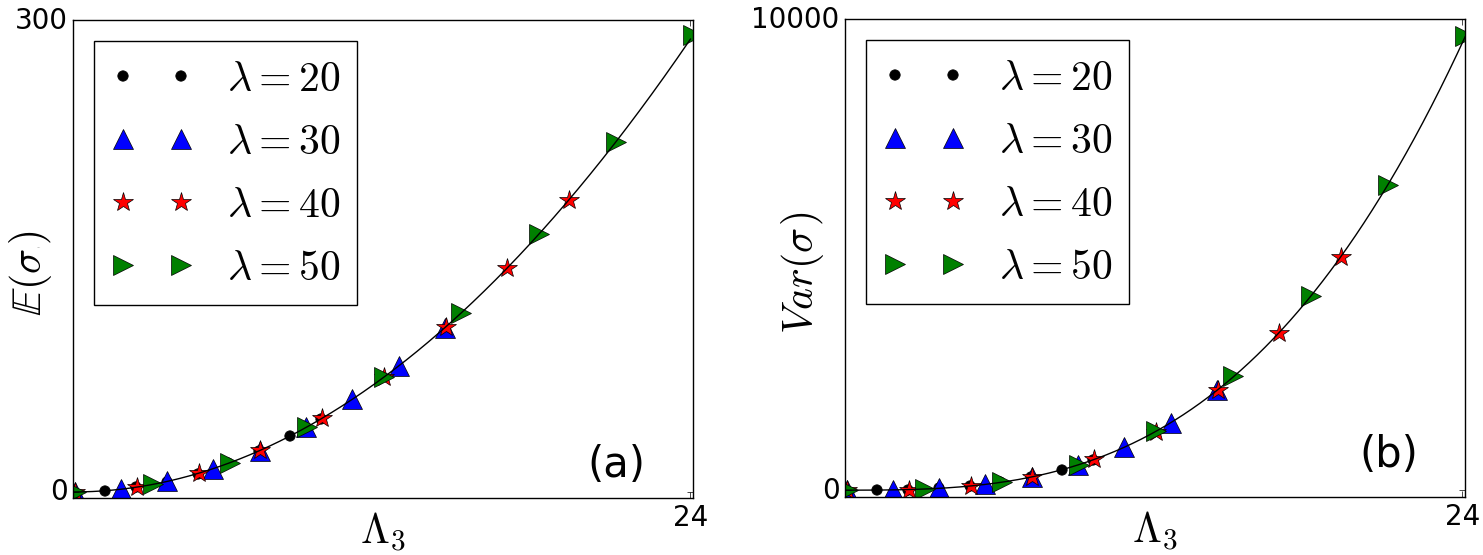} 
\caption{Mean (a) and variance (b) of the number of shortest shortest 3-hop paths $\sigma$ as a function of $\Lambda_3=\lambda(3r_0 \!-\! L)$ calculated numerically from ensembles of $10^6$ realisations for a range of values of $\lambda$ and $r_0/L$. Also illustrated is the analytical results of Eq.(\ref{Expectation2}) and Eq(\ref{VarianceSigma}) in (a) and (b) respectfully (grey line).}
\label{fig3}
\end{figure}


\section{Generalisation to k-Hop Shortest Paths}
More generally for $L\in((k-1)r_0, k r_0)$ with integer $k$ there will be $k-1$ lenses of equal width $|L_i|=kr_0-L$. The method of (\ref{Sec:MeanVariance}) can still be used. For general $k$ we have
\es{
\sigma_{k}=\lim_{l \rightarrow \infty}\sum_{q_{k-1}=1}^{l}... \sum_{q_2=1}^{q_3}\sum_{q_1=1}^{q_2}Y_{k-1,q_{k-1}}...Y_{2,q_2} Y_{1,q_1}, 
\label{sigma4hop}
}
where the $Y_{i, q_i}$ are Poisson with mean $\Lambda_k/l$. 
Using that $\sum_{k=1}^n k^{\theta}\!=\!n^{\theta+1}/(\theta+1)+o(n^{\theta+1})$ we can derive the mean 
\es{
\E[\sigma_{k}]=\frac{\Lambda_k^{k-1}}{(k-1)!}
\label{Esigmakhop}
.}
Now, letting $T_q^{(3)}\!=\!Y_{2,q}\sum_{r=1}^{q}Y_{1,r} $, we recursively define 
\es{
T_q^{(n+1)} =  Y_{n, q}  \sum_{r=1}^q T_{r}^{(n)} 
\label{Tqn}
}
By further defining
$
\tau_l^{(n)} \!=\! \sum_{r=1}^{l}T_{r}^{(n)}
$, such that
$
\sigma_{k}\!=\!\lim_{l \rightarrow \infty} \tau_l^{(n)}
$ we can recursively define the expectation of $\tau_l^{(n)}$ 
\es{
\E[\tau_l^{(n+1)} ]=\frac{\Lambda}{l}\sum_{r=1}^{l}\E(\tau_l^{(n)}  ),
\label{expTrec}
}
where $\Lambda/l$ is the mean of the Poisson variables  $Y_{i,j}$.
Similarly, for the variance of $\tau_l^{(n+1)}$ we calculate
\es{
\sum_{r=1}^l \Var(T_{r}^{(n+1)}) + 2 \sum_{s = 1}^{l-1} \sum_{t=s+1}^l \Cov(T_{s}^{(n+1)}, T_{t}^{(n+1)}).
\label{varsumTrec}
}
Using the recurrence relation we have
\es{
\Var(T_{r}^{(n+1)}) =\left(\frac{\Lambda^2}{l^2}+ \frac{\Lambda}{l}\right)\Var(\tau_r^{(n)})+\frac{\Lambda}{l}\E(\tau_r^{(n)})^2
\label{varTrec}
}
For the covariance we have 
\es{
&\Cov(T_{s}^{(n+1)}, T_{t}^{(n+1)})\! =\! \E[T_{s}^{(n+1)} T_{t}^{(n+1)}]-\E[T_{s}^{(n)}]\E[T_{t}^{(n)}] \\
&\!=\!\frac{\Lambda^2}{l^2} \Big(\E\Big[   \sum_{p=1}^s T_{p}^{(n)} \Big(\sum_{p=1}^s T_{p}^{(n)} \!+\!\!\! \sum_{p=s+1}^t \!\! T_{t}^{(n)}\Big) \Big]\!-\! \E[\tau_s^{(n)}] \E[ \tau_t^{(n)}] \Big)\\
&\!=\!\frac{\Lambda^2}{l^2} \Big(  \sum_{p=1}^s \sum_{r=s+1}^t \! \E[ T_{p}^{(n)}T_{r}^{(n)}]  \!+\! \E[(\tau_s^{(n)})^2]-\E[\tau_s^{(n)} ] \E[ \tau_t^{(n)}] \Big)
\label{Eq:Covrecur}
}
where we have used that $t\!>\!s$ and split the sum into parts. We now have 
\es{
&\Cov(T_{s}^{(n+1)}, T_{t}^{(n+1)}) \!=\!
 \frac{\Lambda^2}{l^2} \Big[ \Var(\tau_s^{(n)}  ) \!-\! \E[ \tau_s^{(n)}] \E[ \tau_t^{(n)}]\\
&+\! \E[\tau_s^{(n)}]^2 \!+\!  \sum_{p=1}^s \! \sum_{r=s+1}^t \! \Cov\left(T_{p}^{(n)},T_{r}^{(n)}\right) \!+\! \frac{\Lambda^2}{l^2}\E[\tau_p^{(n)}]\E[\tau_r^{(n)} ] \Big]
\label{Eq:CovRecurr}
}
Letting $n=k$ and $\Lambda/l= \Lambda_k/l$ we can combine (\ref{expTrec}), (\ref{varsumTrec}), (\ref{varTrec}) and (\ref{Eq:CovRecurr}) to obtain the variance for $\sigma_{k}$. 
For example, for $k\!=\!4$ we have 
$
\Var(\sigma_{4}) = \frac{6\Lambda_4^5 + 15 \Lambda_4^4 + 10\Lambda_4^3}{60}
$.

This recursion relation allows us to derive the variance of $\sigma_k$, which involves evaluating a $(k-1)$-fold sum of products of random variables (see (\ref{sigma4hop})) in terms of a simpler $(k-2)$-fold sum. 


\section{Conclusion \label{sec:conclusion}}
Motivated by the multihop diffusion of information in VANETs, realised through the periodic broadcasts of BSMs as mandated by the DSRC standard \cite{kenney2011dedicated}, we have studied the statistics of the number $\sigma_{k}$ of shortest $k$-hop paths in 1D random networks.
Namely, we have derived simple closed form expressions for the mean and variance of $\sigma_{k}$ for $k=3,4$, provided a recursive formula for general $k$, and have confirmed them numerically using Monte Carlo simulations (see Fig. \ref{fig3}).
We argue that knowledge of such statistics can be used to enhance throughput \cite{rossi2015density}, validate threat events, protect against collusion attacks, infer location information, and also limit redundant broadcasts thus reducing interference.
As an example, consider the realistic scenario where there are about $\lambda \!=\! 100$ vehicles per km, transmission range is $r_0\!=\!0.3$ km, and a vehicle detects an event and broadcasts a BSM containing relevant safety information which should reach at least a range of $L\!=\!1$ km from the epicentre of the detected event.
It follows that the length of the shortest multihop path is $k\!=\! \lceil \frac{L}{r_0} \rceil \!=\!4 $, and that the expected number of shortest paths is $\E[\sigma_{4}] \!=\! 1333.33$ (in either forward or backward direction).
This is clearly unnecessary, and only a fraction of $\nu\in(0,1]$ vehicles should re-broadcast the original BSM.
Thus inverting \eqref{Esigmakhop}, we can calculate the re-broadcast probability $\nu\!=\! \frac{(\varsigma (k\!-\!1)!)^{1/(k-1)}}{\lambda (k r_0\!-\! L)} $, where $\varsigma$ is the target number  of shortest paths, e.g., setting $\varsigma\!=\!10$ we estimate that just $19.5\%$ of vehicles should re-broadcast the original BSM.

\section*{Acknowledgements}
The authors would like to thank the Directors of the Toshiba Telecommunications Research Laboratory for their support.
This work was supported by the EPSRC grant number EP/N002458/1 for the project Spatially Embedded Networks. 


\end{document}